# Quantitative elemental imaging in eukaryotic algae

Stefan Schmollinger[1,2,*], Si Chen[3] and Sabeeha S. Merchant[1,2]

[1]California Institute for Quantitative Biosciences (QB3), University of California, Berkeley, CA 94720

[2]Departments of Molecular and Cell Biology and Plant and Microbial Biology, Berkeley, CA 94720

[3]3X-ray Science Division, Argonne National Laboratory, Lemont, IL 60439

[*] Present address: MSU-DOE Plant Research Laboratory, Michigan State University, East Lansing, MI 48824

Corresponding author: Stefan Schmollinger

612 Wilson Road, MSU-DOE Plant Research Laboratory, Michigan State University, East Lansing, MI 48824, USA, +1 (310) 779-0687, schmolli@msu.edu

Stefan Schmollinger         (https://orcid.org/0000-0002-7487-8014)
Si Chen                     (https://orcid.org/0000-0001-6619-2699)
Sabeeha S. Merchant         (https://orcid.org/0000-0002-2594-509X)

**Abstract**

All organisms, fundamentally, are made from the same raw material, namely the elements of the periodic table. Biochemical diversity is achieved with how these elements are utilized, for what purpose and in which physical location. Determining elemental distributions, especially those of trace elements that facilitate metabolism as cofactors in the active centers of essential enzymes, can determine the state of metabolism, the nutritional status or the developmental stage of an organism. Photosynthetic eukaryotes, especially algae, are excellent subjects for quantitative analysis of elemental distribution. These microbes utilize unique metabolic pathways that require various trace nutrients at their core to enable its operation. Photosynthetic microbes also have important environmental roles as primary producers in habitats with limited nutrient supply or toxin contaminations. Accordingly, photosynthetic eukaryotes are of great interest for biotechnological exploitation, carbon sequestration and bioremediation, with many of the applications involving various trace elements and consequently affecting their quota and intracellular distribution. A number of diverse applications were developed for elemental imaging allowing subcellular resolution, with X-ray fluorescence microscopy (XFM) being at the forefront, enabling quantitative descriptions of intact cells in a non-destructive method. This Tutorial Review summarizes the workflow of a quantitative, single-cell elemental distribution analysis of a eukaryotic alga using XFM.

The aim of this review is to (1) highlight the contributions of different elements to photosynthetic life and the concepts of how organisms control their elemental composition, (2) introduce the methodologies involved in studying elemental distributions in cells, especially X-ray fluorescence microscopy (XFM, XRF), (3) review the current state of XFM studies in eukaryotic algae, and (4) to extract a methodology frame-work for conducting XFM studies from these works. It is our goal to facilitate the entry into the field of elemental research for algae scholars encountering questions of metal homeostasis and elemental heterogeneity for the first time, and to encourage the use of quantitative elemental imaging approaches for the purpose of determining biological function.

**Elemental composition of cells**

The elements of the periodic table are the indivisible foundation of all matter, including all biological life of our planet (Figure 1). Every component of a cell is assembled from a selection of elements, most prominently C, H, N, O, P, S, which are essential and constitute the backbone of proteins, carbohydrates, nucleic acids and lipids (1-3). Essentiality of an element is defined by its irreplaceability and its requirement in metabolism to complete the vegetative or reproductive life cycle (4), while beneficial elements only improve the organisms fitness. The set of essential elements therefore can vary between different organisms, depending on the organism's environmental niche and its required enzyme portfolio. It is estimated that ~ 40 % of all enzymes utilize a uniquely suited element outside the group of macronutrients (CHNOPS) within their catalytic centers to enabling catalysis (5, 6). Most organisms employ several cations (K, Mg, Ca and to a lesser degree Na) and the Cl anion, all of which are abundant constituents of biological matter, to regulate osmotic pressure and pH, build gradients across membranes that facilitate energy production, transport processes or signal transduction, or serve as cofactors or allosteric regulators in metabolites or proteins (7). And lastly, organisms require various additional sets of elements in trace amounts (including many metal ions), to enable chemical functionalities that are not provided from metabolites or amino acids (3, 8). These micronutrients/trace elements are required co-factors for metabolically-essential enzymes, or for enzymes that enable new biochemical capabilities, and consequently their acquisition, intracellular distribution and utilization are a crucial aspect of cellular metabolism (3). Trace elements typically present the most interesting targets for elemental imaging, because of the impact of their chemistry on cell health and metabolism, the range of their abundances in an organism and the dynamic regulation involved in their utilization (3, 9). A set of trace elements is essential for most organisms, including Fe, Cu, Co, Ni, Mn, Zn, Se, V, I and Mo. Redox biochemistry is a central aspect of trace metal utilization. Mn, Cu, and Fe, are therefore among the most abundant and important trace elements for all organisms; at lower abundance Ni, Co and Mo are also required as cofactors by many organisms (5, 10-17). Fe, Cu and Mn, among many important contributions, are critical for photosynthetic electron transfer (18). Ni is used in a wide range of organisms, for example in ureases and hydrogenases (19). Mo, outside of bacterial Mo-nitrogenase, is usually found in the Molybdenum cofactor Moco, which is used in many enzymes, most prominently nitrate reductase (17). Zn is similarly abundant and widespread as a trace element as Fe, Cu and Mn, but is used as a Lewis acid and a structural component for proteins in most organisms (20-22). Se, most prominently, is utilized as selenocysteine in specific enzymes requiring the element in their catalytic centers, for example glutathione peroxidases (23-26). Co is used as a cofactor in a few enzymes directly, but most famously is at the center of cobalamin (also known as vitamin $B_{12}$), which is critical to nitrogen-fixing bacteria (27, 28). V is used in vanadium-dependent nitrogenases and haloperoxidases, and has an important role as an electron acceptor in bacterial respiration (29). I is used in thyroid hormones in vertebrates and haloperoxidases in algae (30, 31).

Othertrace elements (B, Si, As, Br, Sr, Cd, Ba, W, Hg, Pb, La, Ce and Nd) are either employed in very specialized roles in select organisms in a specific environmental niche, or in individual enzymes with a specific beneficial, but non-essential function (32-35).

Many of the trace elements, among other bioactive elements, can also accumulate in organisms involuntarily, using uptake routes for intended cofactors, and interfere with biological processes either in an advantageous, or more commonly a detrimental way.

Therefore, utilization of, especially the redox-active trace elements, comes with inherent risks. The chemical reactivities that make these elements useful in the first place must be controlled intracellularly to avoid unintended reactions (36-38). The concentration of many of these elements in the environment of cells is a critical parameter, determining if the organism is starving for the element as a nutrient, if either abundance or bioavailability is low, or if cell health is threatened by overexposure (39). The toxicity can either be directly attributed to the detrimental reactivities of the elements when uncontrolled, to the production of secondary toxic products, for example reactive oxygen species, or via enzyme mis-metalation. Mis-metalation is largely attributed to the inherent flexibility in proteins and the similar physical properties (ionic radii, charge and coordination preferences) of the biologically common trace metal (40, 41). Most enzymes are tuned to function with a specific metal cofactor. Binding of a different, similar metal at the active site can result in loss-of-function, or worse, the production of unintended products or promotion of side-reactions (38, 42, 43). All organisms therefore carefully control their elemental composition at the point of uptake, resulting in specific cellular quotas, especially in the case of redox active trace elements. Cells also employ elaborate strategies to avoid mis-metalation intracellularly, including the compartmentalization of specific elements to ensure that the correct metal binds to newly-synthesized proteins, or the use of metallochaperones to ensure correct delivery through protein-protein interactions (44-46). Some metals are associated with organic groups or build into large clusters (for example Fe in heme and Fe-S clusters) for similar reasons.

The pathways for trace element metabolism are among the most ancient in biology (47), and the general concepts involved in trace metal utilization are well conserved across organisms. Photosynthetic organisms specifically have unique requirements with respect to the elemental composition, because of the metabolic demand of the photosynthetic apparatus and specific

**Figure 1: Elements of interest to life**

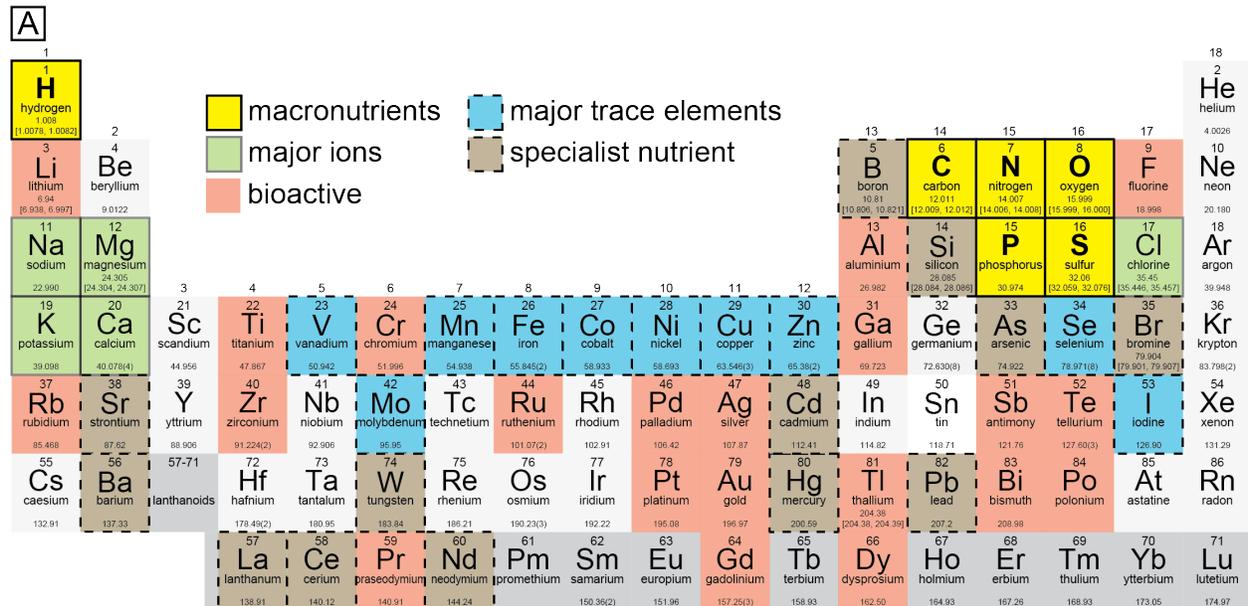

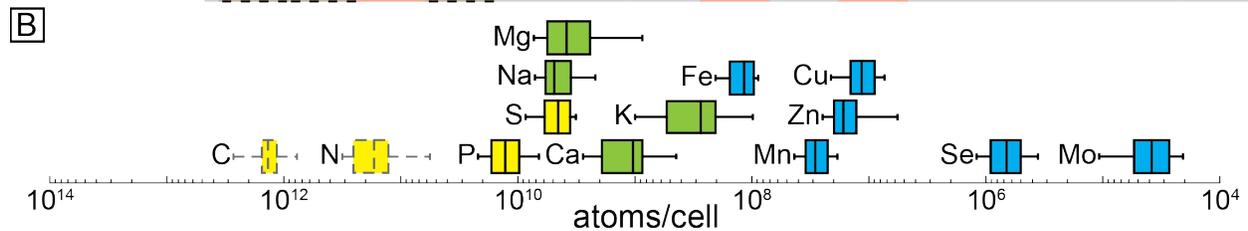

(A) Periodic table of the elements showing relevant elements to living organisms. Elements are grouped into macronutrients (CHNOPS, yellow background), the major cations (Na, Mg, K and Ca, green background), major trace elements relevant for enzymatic function in many organisms (blue background), nutrients that a relevant for specialist's enzyme function (brown background), and elements that affect metabolism (beneficial or toxic) in living organism without being directly utilized in enzymes (bioactive, red background). (B) Elemental composition of the eukaryotic green algae *Chlamydomonas reinhardtii*. Solid lines indicate ICP-MS/MS analysis, dotted lines indicate TOC/TN measurements.

pressures they experience from the environmental niches they occupy. Processes realized by abundant enzymes weigh harder on the specific elemental quotas, and proteins involved in photosynthetic carbon fixation (most prominently photosystem I and II, the Cyt $b_6f$ complex and the enzymes of the Calvin-Benson cycle) are amongst the most abundant proteins in photosynthetic organisms. ~Magnesium is especially important here, with 25% of Mg in plants being found in chloroplasts, where it is integral for the light-capturing Chlorophylls (Mg-containing tetrapyrroles) and in the activation of the carbon-capturing RuBisCO (48–50). The photosynthetic electron transfer chain also requires large amounts of Fe, 80-90% of Fe in leafs is found in chloroplast (51).Most abundantly, Fe is found in photosystem I, which is utilizing three 4Fe-4S centers to transport electrons from the acceptor to the donor side, and in ferredoxin (2Fe-2S center), the soluble carrier protein distributing photosynthetically-derived electrons to various chloroplast processes, including NADPH production, nitrogen assimilation and Chlorophyll biosynthesis (52, 53). Cu in plastocyanin, the soluble electron carrier between the Cyt $b_6f$ complex and photosystem I, and Mn and Ca, in the oxygen evolving complex of photosystem II, are also vital in photosynthetic electron transport (12, 54, 55).

**Intracellular elemental heterogeneity**

The elemental composition of an organism is a dynamic function of the metabolic needs of a cell. As mentioned above, it varies between organisms depending on the functions they employ; but even within a given species it is adjusted between different metabolic or developmental stages, tissues and cell types, and in response to the availability of the nutrients and other stimuli in the environment.

Nutrient limitation is a major driver of acclimation responses to the elemental composition. While limitation to essential elements often results in cell cycle arrest or even cell death, the elemental composition of non-essential, but beneficial trace nutrients can be most variable. Fe bioavailability, for example, is low both in aquatic and cropland environments, mainly because of its low solubility in the more oxidized, but most prevalent, Fe(III) state (56, 57). Its central role in the abundant photosynthetic apparatus, which most primary producers depend on for carbon fixation, assures that Fe availability limits virtually all forms of life (58). An evolutionary adaptation in some photosynthetic organisms to low Fe environments was therefore to use a Cu-containing protein, plastocyanin, instead of the Fe-containing Cyt $c_6$, for the same function, namely, the transport of electrons between Cyt $b_6f$ and photosystem 1 (55, 59). While this reduces the organisms' Fe quota, simultaneously, its dependence on Cu becomes greater. Other acclimation mechanisms also involve the intracellular recycling of metal cofactors and subsequent redistribution to other process, according to a hierarchy of essentiality in the organism (60-63). On the other end of the spectrum, over-accumulation of various elements beyond the necessary quota can affect cell health and occur in polluted or otherwise nutrient-imbalanced environments. Plants, for example, accumulate P when Zn is limiting and vice versa (64-66), while the green alga Chlamydomonas accumulates large amounts of Cu in Zn limiting conditions (67, 68). Sequestration in specialized, intracellular compartments is a common strategy to either detoxify over-accumulating, biologically undesired elements, store scarce nutrients in preparation for periods of limitation or for future generations, sequester a resource away from competitors, or to buffer nutrients temporarily during metabolic transitions (69-72). Acidocalcisomes are lysosome-related organelles, first identified in trypanosomes but widely present in eukaryotes (73). Acidocalcisomes are rich in Ca and P (74, 75), in some organisms also K (76), and can temporarily accumulate various micronutrient metals and toxic elements. The eukaryotic green alga Chlamydomonas, for example, has been found to sequester the trace nutrients Cu, Mn and Fe and the heavy metal Cd in acidocalcisomes in periods of over-accumulation (67, 77-80) . Vacuoles in general are the most important storage sites in eukaryotes, including photosynthetic organisms (70, 72, 81), but other means of storage and sequestration can also be utilized. Proteins like ferritin, a soluble 24-subunit oligomer, can sequester trace elements, ~4500 Fe ions in ferritins case (82, 83), and other organelles like the starch-separated pyrenoid in the chloroplast have been found to contribute to Cd sequestration (79, 80, 84).

Research of metal homeostasis not only aims to identify acclimation processes in natural settings, but can also be exploited in bioremediation or biofortification applications. While heavy metals, $z > 20$, density $> 5$ g/cm$^3$ (85), can be naturally present in specific soil or aquatic environments, more importantly human activities, from mining and industrial production processes to domestic and agricultural practices, have led to an increased contamination of natural habitats with toxic metals (86-88). Cd, As, Pb, Cr and Hg are thereby the most prominent elements, all presenting severe dangers to human health (86). While conventional methods, like chemical precipitation, reverse osmosis, adsorption or electrodeposition are used to remove heavy metals from environments, the use of biological organisms can be much more efficient and cost-effective (89, 90). Photosynthetic algae for example are uniquely positioned to being utilized in the removal of heavy metals in soil and aquatic environments, and research into their trace metal metabolism can greatly facilitate the effectiveness of these processes (91-93). Biofortification on the other hand, is a biotechnological process intended to increase the nutritional value of human nutrition, with major crop plants being generally poor sources of micronutrients (94). Targeting photosynthetic organisms like crops or algae, which are used as animal feed stocks or in natural supplements, is an efficient way to improve nutritional deficiencies (95).

**Quantitative, intracellular elemental distributions**

Whether it is for research or in the pursuit of biotechnological applications, the acquisition of spatially-resolved, quantitative elemental maps is a key tool for researchers to identify the molecular mechanisms involved in elemental homeostasis. Elemental distribution can identify the function, specificity and directionality of membrane transporters, which are involved in the uptake, removal or intracellular distribution of specific nutrients in the cell. Transport mechanisms are equally involved in storage or sequestration efforts. Other proteins involved in elemental homeostasis, for example transcriptional/translational control factors, signal transduction components, chaperones, buffering / protective proteins or metabolites and major client proteins, for example abundant enzymes requiring a specific elemental cofactor, are also required to achieve native elemental distribution. Using specific mutants and analyzing differential elemental distribution maps can facilitating the identification of the function of these proteins in the first place (96). Detailed elemental distribution maps in different stages of nutrition for individual elements or upon other environmental perturbations (pH, temperature, light), can help in identifying the molecular mechanisms utilized in acclimation.

Therefore, over the past decades, a number of analytical techniques have been developed to determine intracellular elemental distribution, utilizing different properties to distinguish the elements. The various techniques all have different advantages and disadvantages, making the different applications quite complementary, especially with regards to sensitivity (detection limit and quantifiable range), obtainable spatial resolution, range of elements that can be (simultaneously) recorded, sample preparation, preservation and the amount of material required, and the kind of artefacts produced either from sample preparation or from the methodology itself. For detailed reviews of the individual methodologies see (97-99), and an in depth discussion of strengths and weaknesses between the most common individual techniques can be found here (100). Two major categories can be distinguished, i) fluorescence microscopy-based techniques using element-sensitive dyes or genetically encoded metal-binding sensors and ii) scanning technologies using either mass-spectrometry (Laser Ablation Inductively Coupled Plasma Mass Spectrometry, LA-ICP-MS; nanoscale Secondary Ion Mass Spectrometry, nanoSIMS) or the detection of element-specific energy signatures (Synchrotron Radiation X-Ray Fluorescence Microscopy, abbreviated either XFM, XRF, SXRF or SRXRF; Particle Induced X-ray Emission, PIXE; Energy Dispersive X-ray Spectroscopy, abbreviated either EDX, EDS or EDXS) to determine the elemental composition (97).

The fluorescence microscopy-based techniques rely on the chemical properties of the individual molecular probe to identify a specific element, sometimes even a specific oxidation state, resulting in a fluorescence change (reversible or irreversible). Coupled with an adequate fluorescence microscope with the necessary resolution, which are available to many laboratories, the probes allow to quickly analyze intracellular elemental distributions in many cells (101). The analysis is limited to a set of compatible, non-overlapping fluorescent signals at a time, but the probes can be utilized in living cells, to specifically assess the labile, accessibility fraction of the element in the cell. Quantitation of the labile metal pool using fluorescence-based probes is possible (102), but access to particularly tightly bound, or less-accessible, sequestered elemental cofactors might preclude capturing the total metal distribution using probes.

Outside of the microscopy techniques, XFM is a popular choice to determine quantitative, intracellular elemental maps (98, 103). XFM takes advantage of the unique electron orbital configurations of each element (Figure 2). Using a highly focus X-ray beam at an energy above the binding energy of a core electron of an element, can result in the removal of the electron from the atom. Upon removal, another electron from an outer shell can rapidly transition to the inner shell and fill the hole. The energy-difference between the orbitals, specific for each transition (Figure 2B), can be emitted as X-ray fluorescence, which can be recorded and analyzed using an energy-dispersive detector (104). Using a highly focused incident X-ray beam and a precisely controllable sample stage, 2D-projection images of the elemental composition of the material in the path of the beam are then assembled spot by spot at high resolution. The low abundances of trace elements in biological samples, at this point of development of the technique, requires the use of highly brilliant synchrotron radiation sources to fast determine subcellular distribution maps, crucially limiting the amount of cells that can be analyzed (99). Benchtop systems however continue to improve (105), and can present an attractive route to improve access to elemental imaging in the future. There are more than 50 synchrotrons globally, many of them offering XFM capabilities; pixel sizes below 100 nm can be achieved at multiple beamlines at synchrotrons including (100, 106-114), which is well below the threshold of what is generally considered subcellular resolution (<1 μm, nanometer scale). Due to the high penetration of hard x-ray photons, XFM can be used to examine much thicker samples compared to those used for electron microscopy, removing the requirement for sectioning and therefore facilitating sample processing and whole-cell analysis simultaneously.

Mass spectrometry-based techniques (LA-ICP-MS or nanoscale SIMS) are alternative options, offering isotope-specific detection of elements (115). In principle, these techniques use a highly focused laser (LA-ICP-MS) or ion beam (SIMS) to release material from a specific spot of a sample (cell section in the case of intracellular studies), which is then analyzed in a mass spectrometer for its elemental composition. This has several advantages over XFM. By using the mass of an element instead of the electron configuration for detection, individual isotopes of element can be distinguished, allowing time course studies with specific isotopes in pulse-chase mode. Lower resolution images of regions of the sectioned material allows to analyze a larger number of cells than can be analyzed with XFM. Both MS-based applications are considered destructive, in the sense that material of the specimen is used for the analysis, compared to the techniques that probe electron configurations or molecular probes, which are generally considered non-destructive, not consuming material for identification of an element. The radiation however used for both of these techniques can, nevertheless, induce changes in the sample during detection. Radiation damage to cell structures, especially with long dwell times or after repeated scans is a concern (116). The resolution of LA-ICP-MS was not previously suited to identify detailed intracellular structures, but submicron resolution has recently been achieved (117), which is a promising development. SIMS at nanoscale can achieve similar resolutions to XFM, sub 100 nm pixel size, but matrix effects have long been a challenge for quantitative aspects. $Cs^+$ and $O^-$ ion beams have a high ionization potential, allowing the determination of intracellular elemental maps, even of elements present at trace amounts (67, 98, 115). The inherent nature of the mass-spectrometry applications, the removal of material for elemental analysis from the surface of cell/tissue sections which can be uneven or inconsistent makes it difficult, but not impossible, to analyze whole cells quantitatively (118).

**Figure 2: Principle of X-ray fluorescence emission**

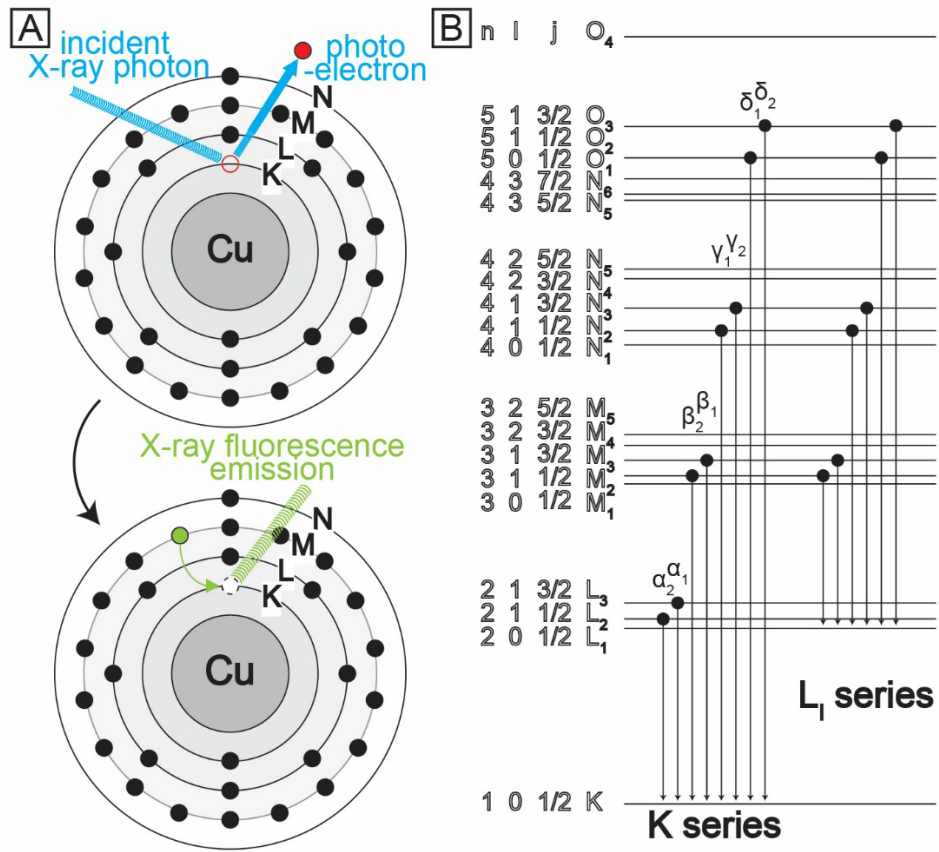

(A) Exemplary schematic of the physics at a Cu atom. An incident X-ray photon from a synchrotron source with sufficient energy to excite a core electron, leaving a hole in the process, which is subsequently filled by an electron from a different shell, while emitting fluorescence equivalent to the energy difference. (B) Overview of the nomenclature of different electron transitions. For X-ray fluorescence microscopy the K and L series are the most relevant for detection.

**Eukaryotic algae as subjects**

Eukaryotic algae are excellent subjects for quantitative elemental imaging studies. Algae is a term of convenience, combining several, diverse eukaryotic groups of photosynthetic organisms (119). Many algae are unicellular eukaryotes of a size that is convenient for imaging applications; not too small to require too high of a resolution on the instrumentation end to determine subcellular distribution, but small enough to not require sectioning in applications that do not require it. They are important primary producers with crucial environmental roles, often inhabiting nutrient-limited environments on land or in the oceans, making them interesting research subjects. Green algae are within Viridiplantae, which also contain the land plants. Together with the red algae (rhodophytes) and glaucophytes in the plantae supergroup, these algae are the result of primary endosymbiosis, where their chloroplast arose from a free living cyanobacterium (120). Outside the plant lineage several diverse groups of algae are found: heterokonts (diatoms, brown algae), dinoflagellates, apicomplexans, haptophytes, crypotomonads, euglenoids and chlorarachniophytes are all eukaryotic algae resulting from secondary or tertiary endosymbiosis, receiving their chloroplast from a eukaryotic donor (119, 121). Within these groups several organisms have been analyzed in detail, but probably no individual alga more than the green alga *Chlamydomonas reinhardtii,* and no group more than the diatoms of the genus Thalassiosira.

*C. reinhardtii* is a unicellular green alga that has been widely used as a eukaryotic, photosynthetic reference system. It is a haploid, facultative phototroph with a short generation time (~ 6h in the presence of a reduced carbon source). Its genome has been fully sequenced, and all three genomes can be targeted for modification (122, 123). Chlamydomonas has been widely used for research on algal metabolism, and has served as a resource for commercial applications of algae as sources of biofuels or bioproducts (124, 125). Its common ancestry with land plants, albeit distant (> 1 billion years ago), allows for cross-comparisons to the intensively studied plant systems, and research performed in either to be cross-informative. Yet, Chlamydomonas is less complex (single, uniform cell), contains smaller gene families and less complex gene structures. In terms of elemental composition, the photosynthetic electron transfer chain is virtually identical with that in land plants, and the alga utilizes a broad spectrum of metal cofactors to sustain its photosynthetic, respiratory and metabolic capabilities. In the past decades, studies have been carried out to identify the major transporters involved in nutrient acquisition and distribution (39, 68, 126-130). Mechanisms for nutrient-sparing and recycling have been discovered (55, 131-134), and storage sites were identified (39, 67, 77, 78, 135-137), making this organism particularly useful for elemental analysis studies.

Thalassiosira is a genus of centric diatoms found in diverse marine and freshwater ecosystems, widely recognized for their substantial contribution to global primary productivity (138-140). Their defining feature are the silicified cell walls, consisting of, species-specific, fine-scaled nano-structures that are built intracellularly in a specialized compartment before being exported to assemble the cell wall (141). The genus contains mostly single-celled species, but, especially in marine species, the single cells can be connected via chitin fibrils to "string of beads" colonies (140). The genomes of *T. pseudonana* (142) and *T. oceanica* (143) have been fully sequenced, protocols for genetic manipulation of *T. pseudonana* (144) and *T. weissflogii* (145). Thalassiosira's elemental metabolism has been of particular interest to researchers, especially that of silicon, with regards to the synthesis of its cell wall, and iron, with regards to irons role in limiting the growth of alga in oceanic, high nutrient/low chlorophyll (HNLC) environments, restricting their potential for carbon sequestration (143, 146-149). Algae have been excellent subjects for elemental imaging studies from the beginning; early work developing XFM instrumentation already involved images of diatoms (150). The first elemental distribution maps of eukaryotic algae using fully developed XFM setups were also taken from diatoms, *Thalassiosira weissflogii* and natural isolates, with 0.5 µm step size at the Advanced Photon Source (APS, Argonne, USA) (151, 152). The authors established the quantitative capabilities for Si, Mn, Fe, Ni, and Zn in single cell analysis and determined the detection limits of the technique. The setup was later used by Twining *et al.* to determine the Fe distribution in natural diatom and dinoflagellate isolates upon ocean iron fertilization (153) or in specific oceanic regions (154), Adams *et al.* used it to determine Cu distribution in the diatoms *Phaeodactylum tricornutum* and *Ceratoneis closterium*, as well as in the green alga *Tetraselmis* sp. (155). Nuester *et al.* determined the Fe distribution in a similar setup with improved resolution, 0.2 µm step size, in the diatoms *Thalassiosira pseudonana* and *Thalassiosira weissflogii* (156). Another diatom, *Cyclotella meneghiniania*, was used in a study by de Jonge *et al.*, significantly improving the resolution of XFM tomography to a useful range (<400 nm) for single cells smaller than 10 µm diameter (157). Diaz *et al.* demonstrated the utility of XFM for green alga research, using Chlamydomonas and Chlorella species to demonstrate Fe sequestration in stationary cells (158). Elemental maps of Fe, Zn and K were obtained from frozen hydrated Chlamydomonas cells with < 100 nm spatial resolution during development of the Bionanoprobe at the Advanced Photon Source (106). The alga, together with a different green alga, *Ostreococcus sp.,* was also the subject for demonstrating the utility of ptychography, a coherent diffraction imaging technique that uses multiple overlapping regions of a cell to provide superior spatial resolution. Deng *et al.* demonstrated <20 nm spatial resolution of ultrastructure imaging of frozen-hydrated algae with ptychography, while simultaneously recording fluorescence spectra to determine the intracellular P, Ca, K, S distributions (76, 159). 3D reconstruction of cellular P, Ca, S, Cl and K distributions and ultrastructure from ptychographic tomography using GENFIRE also took advantage of Chlamydomonas (160). Outside of method development, XFM has proven useful for the characterization of intracellular metal sequestrations sites in Chlamydomonas at the APS, and at the European Synchrotron Radiation Facility (ESRF, Grenoble, France). Researchers found Fe, Cu and Mn to be sequestered in cytosolic vacuoles, acidocalcisomes (67, 77, 78), while the heavy metal Cd was found to be localized both in acidocalcisomes and the pyrenoid in the chloroplast (79, 84). Similarly, the green alga *Coccomyxa actinabiotis* was analyzed at 100 nm resolution in studies aimed to identify the mechanism for Co and Ag tolerance (161). Coccolithophores, eukaryotic algae from the haptophytes and renowned for their calcite exoskeleton, were analyzed for elemental distributions up to Sr using XFM at 50 nm resolution in artificial seawater enriched with trace nutrients (162). They were also the subject of an X-ray ptychography tomography study at 30 nm resolution determining their ultrastructure (163).

**Controlling variance in elemental composition of algae**

Successful elemental imaging for elemental imaging starts with a controlled elemental environment during cultivation of the alga. For this reason, a chemically defined medium is superior to a complex medium recipe. For Chlamydomonas, the most popular media are TAP/TP and HS/HSM (+/− acetate as a reduced carbon source) which are both chemically defined (164). Avoiding components like sea water, peptone or yeast extract would be ideal if possible. Both Chlamydomonas media originally used a trace element mixture recipe developed by Hutner (165), which can vary substantially in its content in between batches. Hutners solution was not specifically optimized for the alga; specifically it lacks Se completely, but contains both Co and B, which are not utilized by the alga (166). Instead of the single trace element additive derived from Hutner, a 7-solution trace element suite was developed specifically for *Chlamydomonas reinhardtii*, that changed the composition accordingly and additionally adjusted the concentrations of the other trace elements (especially Zn, but also Fe, Cu and Mn) to better match the algae's metabolic demand (166). The use of controlled, high-purity chemicals for media preparation additionally ensures reproducible conditions. The water, glass and plasticware used for media preparation and cultivation should be low in contaminants. Acid-washing of glassware is recommended, overnight incubation in 6N HCl followed by thorough rinsing in clean water to remove residual HCl (167). Depending on the condition of interest, liquid pre-cultures used for inoculation at a specific cell density, already grown in the elemental condition intended for analysis, additionally improves the reproducibility between experiments, which can be useful if there are long periods in between replicates (for example in between beamline visits).

**Sample preparation for XFM measurements**

The most crucial component of sample preparation is the trade-off between the necessity to deliver the sample material in the required form to the imaging application of choice and to simultaneously preserve the sample material in an unaltered state that reflects the condition that is analyzed (100). Long incubation steps, changes in temperature, necessary concentration steps (centrifugation) or buffer changes to accommodate fixation can potentially alter the state of the sample material on its way from its habitat (or the growth chamber in a laboratory) to the imaging application. For photosynthetic organisms, changes in illumination and aeration during sample preparation are also potentially critical, as they can affect metabolism quickly (168, 169). Metal-sensitive probes compatible with live cell imaging are probably the gold standard in this regard, allowing the researcher to keep cells as close to the state of interest as possible. For XFM, rapid vitrification of cells in liquid ethane (for example using a FEI Vitrobot Mark IV plunge freezer (76, 106, 159, 160) or similar) is minimally invasive, and preserves the cell in a frozen hydrated state. Successful settings for vitrification of Chlamydomonas on the Vitrobot were reported at a temperature between 20-22°C, humidity of 100%, with a blot time 2 s at blot force 0, blot total 1 (76, 159, 160), or a blot time of 3 s at blot force 2, blot total 1, wait and drain time 0 s (78). For this to be effective, the beamline needs to support imaging at low temperatures (106, 109), and the samples to never thaw after the initial freezing event. A major limitation to using vitrification for sample preparation is the availability of cryo-XFM instruments capable of performing subcellular analyses at low temperatures, and the access to specific equipment like plunge freezers can also be prohibitive. Alternatively, chemical fixation (for example using 4% paraformaldehyde) at an early stage of sample preparation is still used and effective, and allows imaging to take place at XFM instruments at room temperature. The elements of interest are crucial in the choice of fixation for any specific study. Chemical fixation, especially using various popular aldehydes, can be problematic for the preservation of the native state especially of highly diffusible ions, for example Na, K and Cl (109, 170). A previous study (78) showed that the intracellular amounts of Fe and Cu however were similar in vitrified and chemical fixed cells, and the distribution patterns were comparable with chemical fixation, albeit not as crisp as in the frozen-hydrated cells. Aldehydes can also alter membrane permeability (170), which might affect intracellular distribution.

In both cases, the cell material needs to be transferred either to film (79) or $Si_3N_4$ windows (76-78, 106, 159, 160), compatible with the downstream beamlines, prior to freeze-plunging or chemical fixation. More elaborate analyses like ptychography or tomography reconstruction might require specific sample holders, or additional sample preparation, which should ideally already be considered at this stage. The $Si_3N_4$ windows can be pretreated with poly-L-Lysine to improve adhesion of alga cells (a single droplet of poly-L-Lysine applied to the window, incubated for 30 min at 37ºC before the remainder of the droplet is removed and the cells are spotted). For quantitative applications, the cells need to be freed from media remnants (78, 79), with brief washing steps (ideally with water as the last step), which has to be done prior to freeze plunging (161), but can be done after chemical fixation, when cells are already spotted on the carrier (78). Washing is crucial, but extended washing can reduce the number of cells on the sample holder. The force by which the washing solution is applied to and removed from the sample holder contribute to the displacement. Using poly-L-Lysine to assist in adhesion, a good starting point for the cell concentration of a motile Chlamydomonas culture to be spotted on a $Si_3N_4$ window was found to be $1x10^7$ cells/ml, spotting between 50 and 100 µl (78). The concentration of the cells on the sample holder is crucial and should be optimized with the sample preparation procedure in place at a light microscope beforehand. If too many cells are spotted, the analysis becomes significantly more complex, as X-rays can penetrate multiple layers of cells and the resulting fluorescence will be reflective of all cellular material in the path of the beam. The identification of individual cells at the beamline becomes also more tedious, curiously, too little cells on the holder will do the same, by increasing the time of search and movements in between the positions of individual cells; a healthy balance is therefore ideal.

**Determining elemental distribution maps with XFM**

Measurement at the individual beamlines will be dominated by the requirements and specifications of the respective instruments. Samples should be in a state, either by fixation or freezing, to withstand the loading and analysis procedures. Calibration of the X-ray fluorescence emission data is required to allow absolute quantitation of the spatially distributed data, the principle and different strategies are nicely summarized in this review (171).. Briefly, calibration is usually done by comparing the intensity of the fluorescence signal emitted from the individual samples to a calibration curve developed from the measurement of standard thin films containing a few elements with known concentration (172). Matrix matching of the standards and samples is very difficult (171), standards used can therefore be quite different in nature to the biological material analyzed. An in parallel analysis of the elemental composition of the bulk sample material provided for the spatial analysis by other means, for example ICP-MS, is therefore of advantage, and can be used at later stages to interrogate the spatial quantitation of the material (78, 161). Additionally, single-cell organisms like Chlamydomonas grown in asynchronous conditions in constant light (the most typical way), will contain a mixture of cells at various stages in the cell cycle (173-176). The elemental composition of cells can vary with the growth stage of the cells, and specific observations with respect to the elemental composition might be tied to an individual stage of the cell cycle. Synchrotron beamline time allocation is a crucial limiting factor, especially for quantitative experiments, with respect to the number of cells that can be analyzed at the required spatial resolution. Nevertheless, similar to other microscopy techniques, a mixture of cells from different developmental stages should be selected, as much as possible, to reduce the risk of conclusions based on a biased subset. The size of a cell can inform on where an individual cell in a batch culture from an asynchronous population currently resides, selecting cells of various sizes is therefore a straightforward way to avoid oversampling of a specific condition. This can either be done prior to beamline visits, using a light microscope and some coordinate system to record the cell position, but also at the beamline. Scans at a lower resolution prior to data collection are routinely used to minimize instrument time by optimizing the focus area of the cell of interest. They also allow for a first assessment of cell size and ensure cell integrity prior to analysis. This selection process, while crucial, can later contribute to the variance in the quantitative data, especially when the data are normalized per cell.

**Data analysis of spatially distributed elemental data**

Analysis of the acquired emission spectra can be divided into two parts: an accurate quantitation of the data coming from the instrument, and a subsequent analysis of the data in between different cells and conditions, using normalization and statistics. For the first part, different options and strategies are discussed in detail in an excellent review (104). In brief, quantification of fluorescence data can already be achieved by simple binning of counts within a certain energy window. While this can be sufficiently accurate, peak-fitting to determine the area in the same energy window, especially when combined with correcting for background levels or overlapping peaks, has been demonstrated to be more accurate and should be preferentially used if available. Software tools (113, 177-179) greatly facilitate all aspects of image analysis including fitting, ROI analyses and co-localizations. A detailed step-by-step protocol on how to perform and evaluate XFM data fitting using the MAPS software package can be found here (180).

For the subsequent quantitative image analysis software tools are also of great assistance, because they allow extraction of the quantified information for each pixel and can summarize it both for the whole image globally and for selected regions of interest. Initially, we recommend dividing the image into two regions, the part of the image that is covered by the cell and the surrounding region (78, 161). This can be done either manually or algorithm-assisted using an abundant element with good signal-to-noise contrast between the cell and the background, for example S. Both regions contain valuable information. The background, within the same image, contains the signal recovered from the sample holder and remnants of the sample preparation process, which should be subtracted from the biological material to obtain the total cellular content for each of the identified elements. XFM results are determined as the amount of element / area, summarizing the total amount of material in the column of the beam. For background correction, the amount / area in the background is subtracted from the amount / area in the area covered by the cell, before the total cellular amount is calculated, from the area covered by the cell. The amounts of elements in the background region should ideally be small compared to the cell image, if there is a substantial amount of elements of interest in the background, additional washing steps should be added to the sample preparation procedure for the next batch; two consecutive washes of the already spotted cells on the sample holder with clean water were sufficient to have less than 1% of elements of interest / area in the background region compared to the region covered by the cell. The background-corrected cellular content obtained this way can directly be compared to alternative means of quantitation (for example ICP-MS). The quantitative data from this analysis can also be used to evaluate technical aspects, like different approaches to fixation (chemically-fixed vs frozen hydrated samples from the same batch), in-between batch variation or different sample preparation strategies.

Comparisons between different intracellular features is achieved by identifying individual organelles from differential phase contrast images or ptychography reconstruction, or by algorithm assisted identification of specific elemental signatures within the cells. In Chlamydomonas, acidocalcisomes, cytosolic vacuoles with a role in trace metal sequestration (Figure 3A), are rich in P and Ca, and potentially also K (67, 77, 78, 159, 181), contractile vacuoles, involved in osmoregulation, are rich in K and Cl (137, 182, 183), pyrenoids, sites of concentrated $CO_2$ fixation in the chloroplast, might have slightly increased S content (79), large starch shields surrounding the pyrenoid might be visible by the absence of individual elements (Figure 3B). Other organelles or cellular structures can have a distinguishable elemental composition only in specific nutritional or developmental stages, or upon genetic changes, which might then also present an opportunity for selection and further analysis. When algorithm-assisted means are used, it is important to ensure that all the identified regions are completely contained within the cell of interest. Remnants from cell lysis or other cells within the same imaging window might complicate the issue. When intracellular regions are compared, additional normalization or statistical evaluations are necessary to account especially for variance of thickness of the material in the way of the beam. Areas in the center of single cells, where thickness is high compared to peripheral regions, will naturally have a higher amount of element in the 2D projection. Several methods to determine thickness experimentally are discussed previously (104), a more direct way to deal with bias is to utilize a strength of XFM and identify a different elemental distribution from the same cell that shows a uniform distribution, for example S (Figure 3C), and use that for normalization of enrichment analysis (78).

**Figure 3: Element distribution identifying organelles and spatial normalization using S**

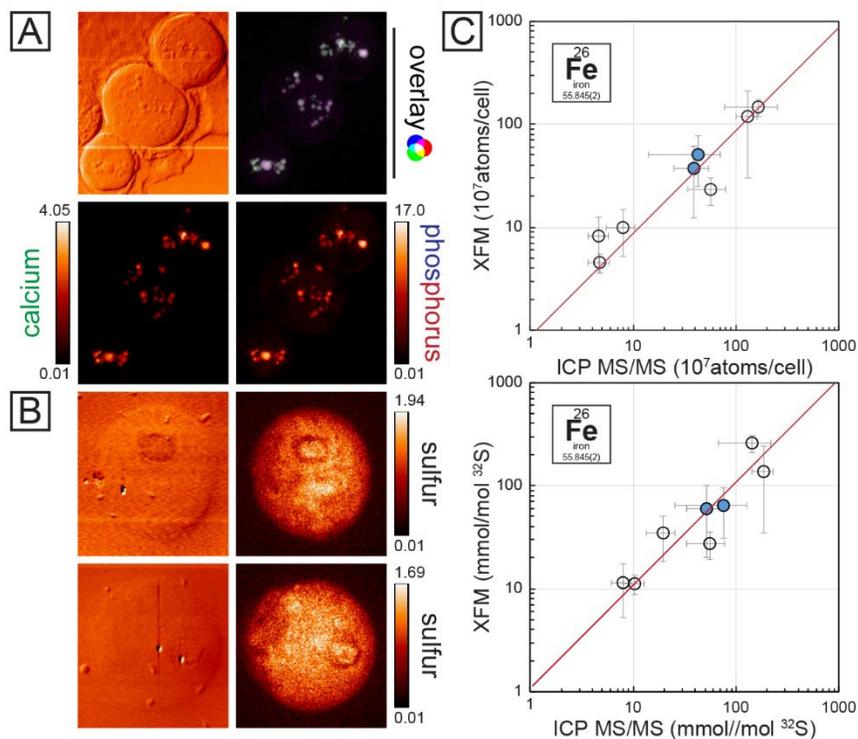

(A) Phosphorus and calcium distribution in Chlamydomonas cells, as well as the overlay, showing acidocalcisomes in Zn deficiency. (B) Sulfur distribution in Chlamydomonas cells, potentially showing starch sheets and pyrenoids. (C) Correlation of Fe content as measured by X-ray fluorescence microscopy (x-axis) and ICP-MS/MS (y-axis), normalized either per cell (top) or using S content (bottom), determined also either with ICP-MS or via XFM. Error bars in x and y direction indicate standard deviation in the measurements between at least 4 individual cells (XFM) or between at least between 3 independent cultures (ICP-MS/MS). Blue fill indicates frozen hydrated samples, grey fill indicates chemical fixation.

**Conclusion**

Eukaryotic algae are crucial primary producers in soil, oceanic and fresh water environments. Their habitats suffer from low trace nutrient bioavailability or heavy metal pollution, sparking research interest into the mechanisms of their metal metabolism. The strategies involved in trace metal distribution can be exploited for biotechnological utilization, either to improve algae growth and consequently their potential for carbon sequestration, for biofortification, or to utilize their mitigation strategies in bioremediation applications. X-ray fluorescence microscopy (XFM, XRF) can be a powerful tool to investigate the intracellular elemental distribution of (trace) nutrients in eukaryotic algae quantitatively, allowing to assign function to individual components involved in managing elemental homeostasis, or to identify acclimation strategies or useful phenotypes. This Tutorial Review summarizes the state of research involving subcellular elemental distributions determined using XFM with eukaryotic algae as subjects, and provides a workflow of a quantitative elemental distribution analysis for eukaryotic alga.